# Temperature-dependent electronic ground state charge transfer in van der Waals heterostructures

Soohyung Park,[1,†] Haiyuan Wang,[2,3,†], Thorsten Schultz,[4,5] Dongguen Shin,[4] Ruslan Ovsyannikov,[5] Marios Zacharias,[2,6] Dmitrii Maksimov,[2,7] Matthias Meissner,[8] Yuri Hasegawa,[8] Takuma Yamaguchi,[8] Satoshi Kera,[8] Areej Aljarb,[9] Mariam Hakami,[9] Lain-Jong Li,[9,10] Vincent Tung,[9] Patrick Amsalem,[4] Mariana Rossi,[2,7,*] and Norbert Koch[4,5,*]

[1] Advanced Analysis Center, Korea Institute of Science and Technology (KIST), Seoul 02792, South Korea

[2] Fritz Haber Institute of the Max Planck Society, 14195 Berlin, Germany

[3] Chaire de simulation à l'échelle atomique (CSEA), Ecole Polytechnique Fédérale de Lausanne (EPFL), CH-1015 Lausanne, Switzerland

[4] Humboldt-Universität zu Berlin, Institut für Physik & IRIS Adlershof, 12489 Berlin, Germany

[5] Helmholtz-Zentrum für Materialien und Energie GmbH, 12489 Berlin, Germany

[6] Department of Mechanical and Materials Science Engineering, Cyprus University of Technology, 3603 Limassol, Cyprus

[7] Max Planck Institute for the Structure and Dynamics of Matter, 22761 Hamburg, Germany

[8] Institute for Molecular Science, 444-8787 Okazaki, Japan

[9] Physical Sciences and Engineering, King Abdullah University of Science and Technology, Thuwal 23955-6900, Saudi Arabia

[10] Department of Mechanical Engineering, The University of Hong Kong, Pok Fu Lam Road, Hong Kong

[†] both authors contributed equally to this work
[*] corresponding authors: rossi@fhi-berlin.mpg.de, norbert.koch@physik.hu-berlin.de

**Abstract**

Electronic charge rearrangement between components of a heterostructure is the fundamental principle to reach the electronic ground state. It is acknowledged that the density of states distribution of the components governs the amount of charge transfer, but a notable dependence on temperature has not yet been considered, particularly for weakly interacting systems. Here, we experimentally observe that the amount of ground state charge transfer in a van der Waals heterostructure formed by monolayer $MoS_2$ sandwiched between graphite and a molecular electron acceptor layer increases by a factor of three when going from 7 K to room temperature. State-of-the-art electronic structure calculations of the full heterostructure that account for nuclear thermal fluctuations reveal intra-component electron-phonon coupling and inter-component electronic coupling as the key factors determining the amount of charge transfer. This conclusion is rationalized by a model applicable to multi-component van der Waals heterostructures.

# Introduction

The discovery that atomically thin layers can exist at room temperature in air marked the launch of research on two-dimensional (2D) materials.[1,2] A single layer of a 2D material may be insulating (e.g., hexagonal boron nitride), semiconducting (e.g., $MoS_2$), or conducting (e.g., graphene), and thus all electrical material properties required for the construction of an electronic or optoelectronic device are available in the monolayer limit.[3-6] Stacks of such monolayers are typically bound by weak van der Waals (vdW) interlayer interactions, and vdW heterostructures have emerged as prime candidates for realizing electronic and optoelectronic functions in the smallest possible volume. The type and efficiency of the functionality that can be achieved depends critically on the electronic energy level alignment across the heterostructure,[7] and substantial charge density re-arrangement upon contact can occur, depending on the electronic structure of each component. Consequently, charge rearrangement and also charge transfer (CT) phenomena that define the electronic ground state of vdW heterostructures must be thoroughly understood for knowledge-guided device design. Beyond layered inorganic 2D materials, organic molecular semiconductors are also attractive as a component in vdW heterostructures, because they extend the range of available energy gap values and feature strong light-matter interaction.[8,9] Furthermore, it has been recognized that strong molecular electron acceptors and donors can be employed as dopants for numerous 2D semiconductors, particularly for transition metal dichalcogenides (TMDCs).[10-14] This doping, on the one hand, allows controlling the Fermi level ($E_F$) position and mobile carrier density in the semiconductor, and, on the other hand, enables manipulation of optical quasi-particles, such as charged excitons (positive or negative trions) in TMDC monolayers.[15-19] Such vdW heterostructures can be ideal model systems for exploring intriguing physical phenomena,[20,21] and they can pave the way for a broader class of device applications.

Despite many opportunities offered by vdW heterostructures, the current understanding of contact-induced charge density re-arrangement phenomena and inter-layer ground state CT is not well established for such weakly bound stacks, mostly due to their challenging complexity. For instance, organic semiconductors feature vast spatial degrees of freedom, and 2D semiconductors experience substantial band structure renormalization as function of the thermodynamic conditions[22,23] and of the surrounding environment, including a supporting substrate.[11] A recent study on vdW heterostructures comprising the molecular electron acceptor 1,3,4,5,7,8-hexafluoro-tetra-cyano-naphthoquinodimethane (F6TCNNQ) deposited on monolayer (ML) $MoS_2$ revealed the key role of the substrate.[11] When insulating sapphire was used as substrate for ML-$MoS_2$, electrons were transferred from *n*-type gap states (induced by native sulfur-vacancies) to molecular acceptors. In contrast, when sapphire was replaced by highly oriented pyrolytic graphite (HOPG), the *n*-type gap states of $MoS_2$ were emptied into the charge reservoir of the conductive substrate and the electron transfer took place directly from HOPG to F6TCNNQ, as schematically shown in Figure 1. This scenario was discussed in analogy to a parallel plate capacitor, where positive and negative charges reside on the two "plates", HOPG and F6TCNNQ, respectively. The ML-MoS$_2$ then represents a dielectric layer experiencing the electric field between the plates, and in this model the frontier energy levels of the semiconductor are not involved in CT. Considering only this picture, the amount of charge transferred across the vdW heterostructure to reach electronic equilibrium is determined by the density of states (DOS) distribution near $E_F$ in HOPG and that of the lowest unoccupied molecular orbital (LUMO) level manifold in F6TCNNQ, as well as their energy offset. However, conjugated molecules exhibit strong electron-vibrational coupling, i.e., the electronic energy levels change upon intramolecular bond length alterations. This translates into a variation of the ground state electronic levels' energy when vibrational modes are populated.[24–26] Therefore, the temperature-dependent population of vibrational modes is expected to alter significantly the electron distribution near $E_F$ and hence the magnitude of CT in vdW heterostructures. This dependence of CT on temperature has not been investigated in vdW heterostructures to date. An indication of the relevance of temperature on CT is the observation that the electrical characteristics of ML-MoS$_2$-based field-effect transistors depend notably on temperature when a molecular acceptor is deposited onto the semiconductor.[14] For molecular acceptors from the tetracarboxylic-dianhydride family deposited on silver single crystals it was found that decreasing the temperature significantly increased the amount of CT, due to the shortening of the molecule-metal distance at lower temperature.[27,28] However, these systems are characterized by covalent interactions and pronounced orbital hybridization, which makes a direct relation to vdW heterostructures uncertain. In fact, for the vdW heterostructure F6TCNNQ/HOPG we find that the amount of CT is virtually independent of temperature (see Supplementary Figure S3).

Here, we demonstrate with angle-resolved photoelectron spectroscopy (ARPES) that the amount of ground state CT in the prototypical vdW heterostructure F6TCNNQ/ML-MoS$_2$/HOPG increases with increasing temperature, gradually and reversibly. This unexpected phenomenon is rationalized *via* high level electronic structure calculations of the full heterostructure, which account for electron-phonon coupling in the adiabatic limit through sampling of stochastic phonon displacements at different temperatures. The calculations show a shift to lower energies and broadening of the acceptor's LUMO manifold with increasing temperature that enhances CT. Within a simple model, it is shown that even in the absence of large-scale structural changes, the energy-gap renormalization of MoS$_2$ and F6TCNNQ with temperature can contribute to the CT enhancement. Therefore, the earlier suggested capacitor model falls short of correctly describing such systems, because it only considers electrostatic interaction.

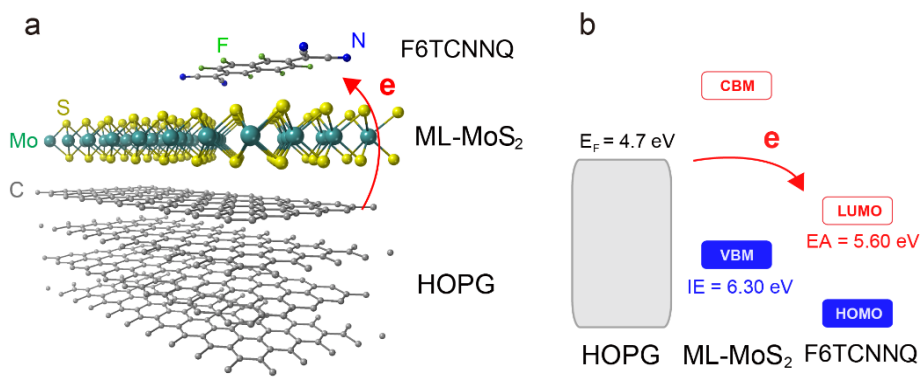

**Figure 1 | Ground state charge transfer in the vdW heterostructure F6TCNNQ/ML-MoS₂/HOPG.** (a) Schematic configuration and (b) energy level diagram of the investigated F6TCNNQ/ML-MoS₂/HOPG van der Waals heterostructure.

## Results

**Temperature-dependent charge transfer in the F6TCNNQ/ML-MoS₂/HOPG vdW heterostructure**

ARPES allows tracking the valence band structure of ML-MoS₂ and the characteristic valence fingerprints of negatively charged F6TCNNQ molecules (anions).[29] For a vdW heterostructure consisting of ca. 0.5 ML F6TCNNQ on ML-MoS₂/HOPG the corresponding energy distribution curves (EDCs) in the near-$E_F$ region from ARPES measurements are shown in Figure 2a, for different temperatures. At room temperature (300 K), this energy region is dominated by two features centered at 0.2 eV and 0.8 eV binding energy (see Supplementary Figure S5), as the valence band onset of MoS₂ is at a binding energy higher than 1.5 eV (see Supplementary Figure S6). These can readily be assigned to emission from the singly occupied former LUMO level of the F6TCNNQ anion (denoted as L*) and its relaxed highest occupied molecular orbital (HOMO) level (denoted as H*), due to electron transfer from HOPG.[11] Spectra recorded at lower temperature exhibit the same features, but with incrementally reduced intensity. We note that the same molecular anion emission features can be identified for F6TCNNQ deposited directly onto HOPG, as shown in Supplementary Figure S3, but no intensity variation as function of temperature was observed. The acceptor molecules adopt a face-on lying orientation on MoS₂ supported by HOPG from room temperature to 77 K, as determined from X-ray absorption measurements (see Supplementary Figure S7), and it is reasonable to assume that this configuration persists even at lower temperatures. Consequently, the area of the emission features L* and H* is representative of the fraction of molecular anions on the surface. Plotting this area ξ, normalized to the signal at 300 K, as function of temperature (Figure 2b), reveals a gradual decrease in the range between room temperature and 7 K. This is a direct indication that the fraction of charged F6TCNNQ molecules at 7 K is only about one third of that found at 300 K. The fraction of charged molecules at 300K was calculated from the transferred electron density divided by the surface molecule density of F6TCNNQ and found to be 21.7 %, while that at 7 K was 7.2 % (for details see Supplementary Section 3). Note that the temperature-dependent change of ξ is fully reversible, as the temperature sequence of this experiment started at 300 K, followed by cooling to 7 K, and subsequent warming-up steps until reaching 300 K again. To provide further support for this finding of smaller CT amount at lower temperature, we analyze the corresponding $MoS_2$ valence band maximum (VBM) positions at the K and Γ points of the ML's Brillouin zone in Figure 2c (selected band structure data plots are shown in Supplementary Figure S6). While the VBM positions of bare $MoS_2$/HOPG show very little change (less than 50 meV) between 300 K and 7 K, they shift by up to 260 meV with F6TCNNQ on top. Furthermore, the trend of VBM shifts as function of temperature follows that already seen for ξ in Figure 2b.

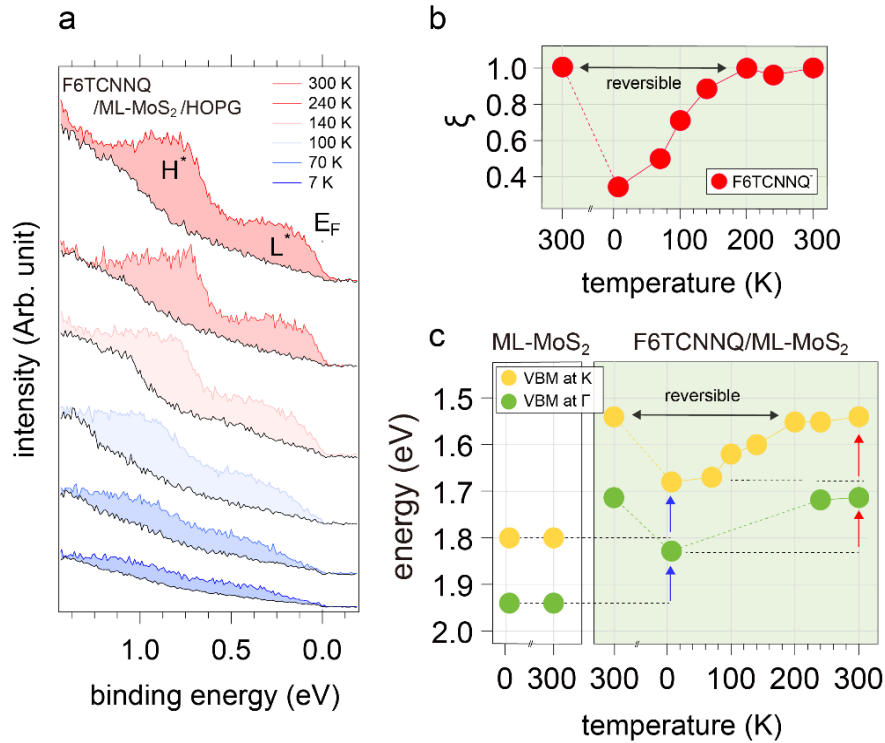

**Figure 2 | Temperature-dependent ground state charge transfer in the vdW heterostructure.** (a) Energy distribution curves (EDCs) of $MoS_2$/HOPG near the Fermi-level as a function of temperature, before (black) and after (colored) 0.5 ML F6TCNNQ deposition. The additional density of states after F6TCNNQ deposition is due to negatively charged acceptors (F6TCNNQ⁻) only. (b) Relative change in charged F6TCNNQ (ξ) as a function of temperature, where the temperature-axis also reflects the experimental sequence, i.e., starting at 300 K, cooling to 7 K, followed by incremental return to 300 K, to highlight the reversibility. (c) Valence band maximum (VBM) of $MoS_2$ at the Γ and K point plotted as a function of temperature with/without absorption of F6TCNNQ. L* and H* denote the partially filled LUMO and relaxed HOMO of charged F6TCNNQ.[11]

These observations are consistent with the overall CT mechanism shown in Figure 1, i.e., electron transfer from HOPG to F6TCNNQ, whereas ML-MoS$_2$ experiences the electric field created by the separated charges and its energy level shift accordingly. The electric field is directly proportional to the amount of CT in the capacitor model, and the same temperature-dependent trend for ξ and the VBM shifts is expected, and verified here by comparison of Figure 2b and 2c. For the vdW heterostructure investigated here, the position of $E_F$ in the semiconductor gap can be tuned by ca. 260 meV simply by varying the temperature, as the different CT amount changes the electric field from approximately 0.26 V/nm at 7 K to 0.57 V/nm at 300 K (estimated with an effective distance between HOPG and the molecular layer of 9 Å). The origin of the temperature-dependent CT amount, however, remains to be elucidated. If lower temperature was to simply decrease nuclear motion and thus reduce the average vertical inter-atomic distances in the heterostructure, the CT amount would be expected to increase with decreasing temperature in the capacitor model; the opposite is observed in experiment. Therefore, to shed light on this observation, elaborate first-principles modeling is employed to obtain more insight into how temperature influences the density of states (DOS) of F6TCNNQ/ML-MoS$_2$/HOPG.

**Charge transfer in vdW heterostructures**

We searched for stable structures of the vdW heterostructure composed of either one (dilute case) or two F6TCNNQ molecules on a 4×8 ML-MoS$_2$ unit cell, which corresponds to almost the same molecular density as in experiments (see SI Section 3). We employed density-functional theory with a range-separated hybrid functional (see computational details in Methods). Globally, the acceptor molecules adopt a flat-lying orientation on ML-MoS$_2$ with a small corrugation of adsorption energy (maximum 150 meV) with respect to lateral displacements (see Supplementary Figure S8 and Table S2). The most stable structures are displayed in Figure 3a and 3d. This lying orientation of the molecules agrees with experiment (see Supplementary Figure S7). Molecules placed on free-standing ML-MoS$_2$ are charge-neutral, as expected from the respective energy levels [*c.f.* experimental values in Figure 1b and calculated projected DOS (PDOS) in Supplementary Figure S9]. Upon adding a commensurate HOPG substrate (modelled by a 5×10 supercell with four layers, totalling 522 atoms in the model) to F6TCNNQ/ML-MoS$_2$, the molecules become negatively charged in this model. This observation is supported by the charge density rearrangement plotted in Figure 3c, and is consistent with the band structures and PDOS shown in Figure 3b, where the zero of energy crosses the PDOS of the molecular LUMO levels. In agreement with the capacitor model discussed above in relation to Figure 1b, the calculated plane-averaged differential charge density (Δρ) plot (Figure 3c) reveals an accumulation of positive charge in HOPG and corresponding negative charge in the molecular layer, giving rise to the electric field drop across ML-MoS$_2$ that shifts its energy levels. The partially occupied LUMO level appears as an almost completely flat band, in Figures 3b and 3e. We confirm that this is the molecular

LUMO by visualizing the orbital in space and comparing it to the orbital of the isolated molecule (see Supplementary Figure S10). We note that the orbital shows non-vanishing hybridization with ML-MoS$_2$. This is a direct manifestation of the finite electronic coupling between the two components, even if their interaction can be classified as van der Waals type.

In the dilute case, the LUMO is partially occupied by 0.3 e, whereas the partially occupied LUMO orbitals of the two non-equivalent molecules in the densely-packed case are not equally charged, i.e., we find that one molecule presents a partial occupation of 0.25 e on its partially filled LUMO and the other 0.1 e. The fractional charge transfer observed here (including the disproportional CT for two molecules in the unit cell) obtained with a range-separated hybrid functional, in contrast to the experimental evidence of integer charge transfer, is likely a consequence of the remaining electron delocalization error present in this functional, allied to the need of much larger supercells that are not computationally feasible to simulate integer CT.[30,31] With this static description (i.e., without temperature-induced effects from nuclear motion) of the vdW heterostructure we obtain adequate qualitative agreement between experiment and theory. In the following, we continue by including quantum zero-point motion and temperature-dependent effects in our calculations.

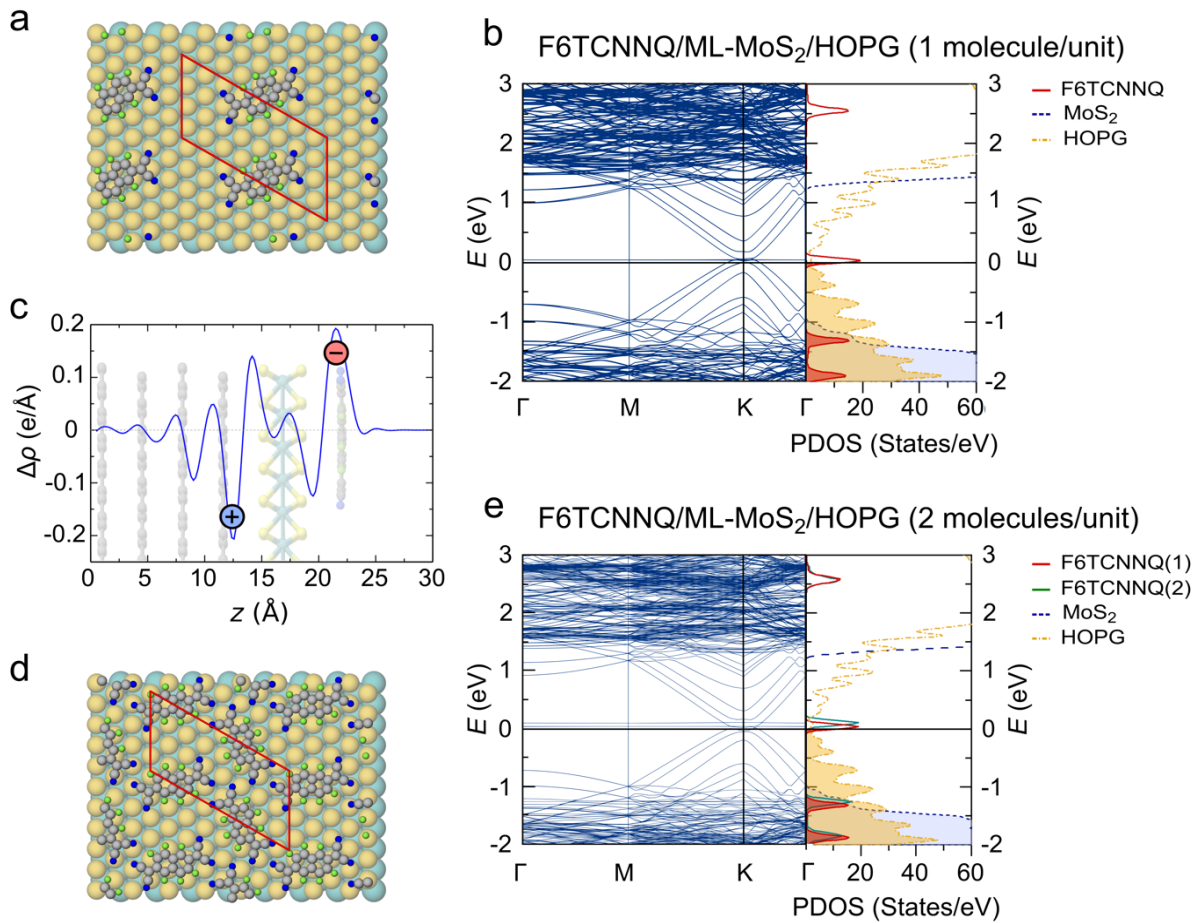

**Figure 3 | Calculated band structure of static F6TCNNQ/ML-MoS$_2$/HOPG heterostructures.** (a) Top view of dilute F6TCNNQ/ML-MoS$_2$(4×8)/HOPG. (b) Band structure and PDOS (DFT-HSE06) corresponding to the supercell shown in panel a. The band structure is folded and PDOS and band structure have been shifted by 5.4 eV, such that the electronic occupation distribution drops below 0.09 above 0 eV (see Methods section for details). (c) Calculated plane average differential charge density ($\Delta\rho = \rho_{\text{F6TCNNQ/MoS}_2\text{/HOPG}} - \rho_{\text{F6TCNNQ}} - \rho_{\text{MoS}_2\text{/HOPG}}$) for the hybrid optimized structure. Positive values mean electron density accumulation (negative charge) while negative values mean electron density depletion (positive charge). (d) Top view of 2 F6TCNNQ/ML-MoS$_2$(4×8)/HOPG. (e) Band structure and PDOS (DFT-HSE06) corresponding to the supercell shown in panel d. The band-structure is folded and PDOS and band structure have been shifted by 5.34 eV, such that the electronic occupation distribution drops below 0.09 above 0 eV.

**Effect of temperature on charge transfer in the vdW heterostructure**

We use the total electronic DOS (TDOS) and the PDOS to analyze the possible impact of finite-temperature nuclear motion on the CT of our vdW heterostructure, for the dilute case. We pick this structure because it is sufficient to grasp the CT, as shown in the previous section. The two aspects that can change the position of electronic states, namely thermal population of vibrational modes and thermal lattice expansion, are investigated separately here. In particular, thermal fluctuations are captured through thermodynamic averages on stochastically sampled nuclear configurations consistent with the distribution of quantum harmonic phonons at different temperatures[32,33] (details given in Methods and Sections 7 and 8 in the SI). The TDOS around the Fermi level in Figure 4a (HSE06 functional) shows a pronounced shift to lower energies between 50 and 150K. It also shows that the occupation of vibrationally excited states according to three different temperatures up to 300 K leads to an obvious broadening compared to the static calculation. The molecular partially-occupied LUMO level dominates the TDOS within this energy window at all temperatures, as corroborated by the PDOS in Figure 4b. The area under this peak integrates to two, because there is no distinction between the two spin channels. Increasing the temperature results in a broadening of the partially occupied LUMO peak and a shift of its maximum towards lower energies (see Figure 4b). This also indicates an increase of occupation of this level, which can be interpreted as an increase of the fraction of molecular anions in the heterostructure. The electronic population of this state has been calculated through integration of the local PDOS with the electronic occupation distribution used in the calculations (see Methods section) and is shown in the inset of Figure 4b. The fraction increases by around 30% from 50 to 300 K. Most of the changes are observed between 50 and 150 K, and the modes that become thermally populated in this temperature range correspond to hindered rotations of the molecules with respect to ML-MoS$_2$ and breathing modes of the MoS$_2$ lattice, both of which influence the adsorption geometry.

The effect of thermal lattice expansion, which is comparably pronounced for MoS$_2$,[34,35] was investigated with static calculations. The results shown in Supplementary Figure S11 show that the effect of the expansion on the electronic orbitals is very small, but nevertheless shifts the partially occupied LUMO to lower energies with increasing temperature. Consequently, nuclear fluctuations and lattice expansion work hand in hand to promote CT in the vdW heterostructure with increasing temperature, as found in our experiment.

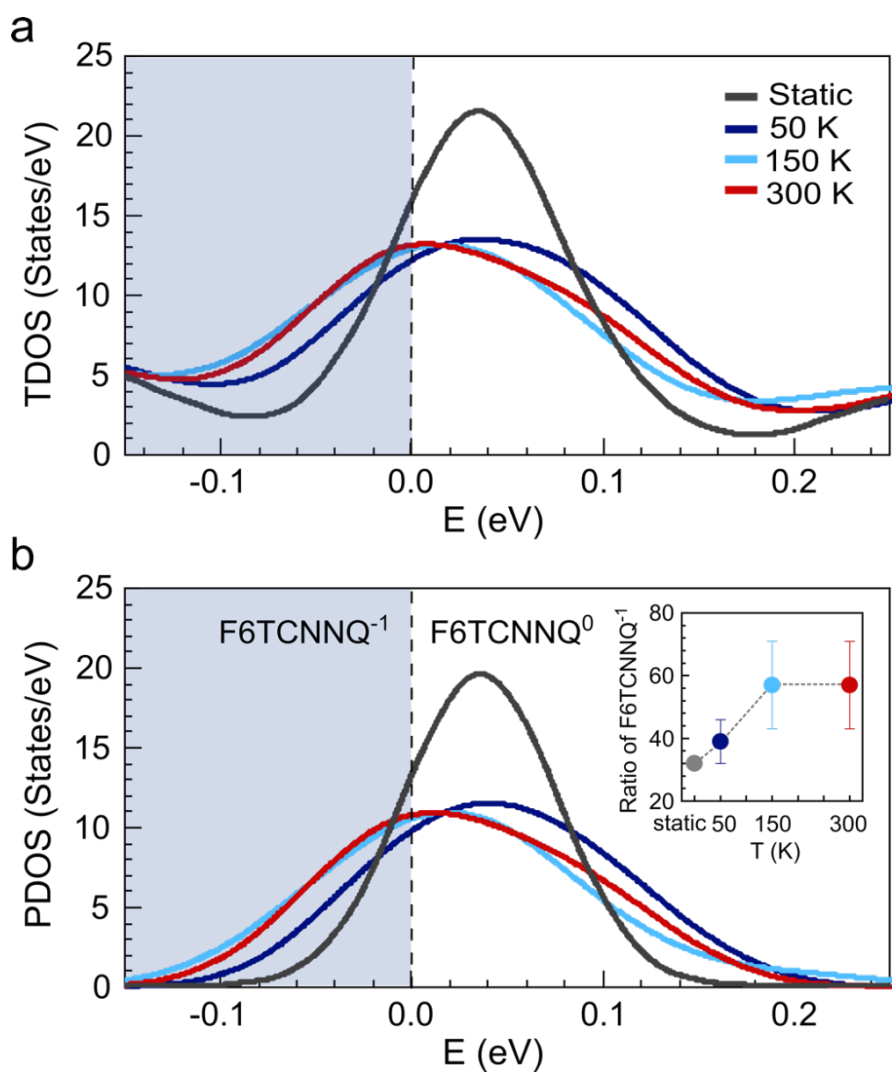

**Figure 4 | Nuclear thermal fluctuations and DOS near Fermi level.** (a) TDOS of the model system of F6TCNNQ/ML-MoS$_2$/HOPG as a function of temperature. (b) PDOS of the model system representing the partially occupied LUMO of F6TCNNQ as a function of temperature. Simulations were performed with occupation smearing (see text). The curves were all rigidly shifted by 5.4 eV, such that the electronic occupation distribution drops below 0.09 above 0 eV. The ratio of charged molecules

(given in %) in the inset of panel (b) were calculated by integration of the local density of states of the LUMO orbital multiplied by the occupation distribution function (see Methods section). The total area under the peaks in panel b integrate to two. Error bars correspond to the remaining statistical uncertainty of the stochastic sampling.

## Discussion

Considering the above results for our specific vdW heterostructure consisting of the molecular electron acceptor F6TCNNQ, the 2D semiconductor $MoS_2$, and the conductor HOPG, one reason for the temperature-dependent amount of charge transfer can be rationalized as consequence of the electron-phonon coupling that is particularly pronounced for ML-$MoS_2$ and the molecules. The molecules' LUMO level, which receives electrons from HOPG and slightly hybridizes with $MoS_2$, is shifted to lower energies and broadened with increasing temperature, resulting in a larger fraction of molecules available for CT. In addition, it is well known that increasing temperature also induces a band-gap renormalization for $MoS_2$[36,37] (see Supplementary Figure S12 and S13). These phenomena, however, cannot be captured by the capacitor model discussed in the introduction, because the amount of CT is expected to depend on the DOS of the electron donor (HOPG) and acceptor (F6TCNNQ) only, and the energy levels of $MoS_2$ in between should play no role. Hence, the electrostatic model neglects a further key aspect. As seen from our calculations, the finite electronic coupling between F6TCNNQ and $MoS_2$, despite the weak vdW type of interaction, also influences the electronic ground state. The electronic coupling between the different parts of the heterostructure, and the temperature renormalization of the $MoS_2$ energy gap and the molecular frontier orbitals, can also impact the character of the ground electronic state and thus the population of the molecular LUMO level. The semiconductor in the middle of the vdW heterostructure thus takes on an active role as "bridge" between electron donor and acceptor.

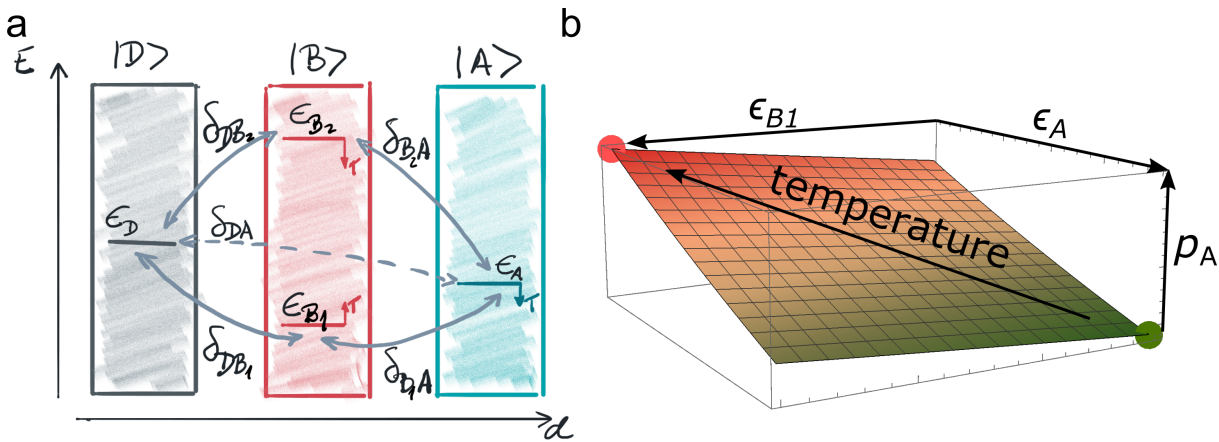

**Figure 5 | Sketch of simple single-particle "donor-bridge-acceptor" model employed in this work.**
(a) Relative energies ($\epsilon$) of electronic levels of HOPG donor (D), $MoS_2$ bridge (B) and F6TCNNQ acceptor (A) systems, schematically reflecting the isolated-system level alignment (see SI). Electronic couplings are denoted by $\delta$. Effect of temperature (T) on the energy levels due to electron-phonon coupling is denoted by arrows. (b) With increasing temperature, the relative position of $\epsilon$ changes due vibronic couplings. The population of the adiabatic state $|A\rangle$ ($p_A$) in the ground state of the interacting system is sensitive to the renormalization of the molecular LUMO level ($|A\rangle$) and the VBM of $MoS_2$ ($|B_1\rangle$) as schematically shown in the figure.

A simple illustration of this effect can be achieved by a single-particle "donor-bridge-acceptor" toy-model. In this model, we denote the donor (D), bridge (B), and acceptor (A), with their respective energy levels $\epsilon_j$ (identified by corresponding subscripts), as well as the electronic couplings between all levels by $\delta_{nm}$ (and corresponding subscripts), as schematically shown in Figure 5a. The resulting Hamiltonian in the adiabatic basis of the isolated components (see in Supplemental Information Section 9) can be constructed by the calculation of $\epsilon_j$ and by assuming reasonable values for the couplings in vdW heterostructures (0.1-0.5 eV) as discussed in more detail in the SI. Solving the Schrödinger equation for this Hamiltonian shows that the ground state contains mainly contributions from the $B_1$ and A states. The thermal population of state A (equivalent to molecular LUMO) in the ground state of the interacting system can be straightforwardly calculated. We can then analyze how this population changes by shifting the position of $MoS_2$ and F6TCNNQ energy levels according to their temperature-induced vibronic renormalization, which we have also accurately calculated for each component in isolation (see Section 8 in the Supplemental Information). The qualitative picture sketched in Figure 5b is found, which changes only quantitatively with different values of couplings in the system. Going along the $\epsilon_A$-axis would correspond to considering only a renormalization of the LUMO level of the molecule with temperature, while moving along the $\epsilon_{B1}$-axis considers only the temperature-dependent VBM renormalization of $MoS_2$. To obtain the full effect of temperature on the ground state CT from D to A, however, necessitates moving along both directions simultaneously, as apparent from the plot. This coupled effect would be absent without the bridge. Although this model is too simple and empirical for a full understanding of the physical processes involved in the real system, it captures the key components that aid CT within vdW heterostructures.

## Conclusions

The amount of ground state charge transfer – to reach electronic equilibrium – in multi-component vdW heterostructures is found to be dependent on the intra-component electron-phonon coupling and the inter-component electronic coupling. Both couplings vary with temperature, and therefore the charge transfer becomes temperature-dependent. Here, this is exemplified for a prototypical vdW heterostructure consisting of graphite/monolayer-$MoS_2$/F6TCNNQ, where the ground state charge transfer amount at room temperature is found to be ca. three times higher than at 7 K. These findings are evidently significant for all studies and applications of vdW heterostructures where temperature is a variable. Furthermore, this insight could be used to bestow electronic and optoelectronic devices with unique temperature-dependent functionalities.

**Methods**

**Sample preparation.** ML-MoS$_2$ were grown on sapphire *via* chemical vapor deposition (CVD) and transferred onto an HOPG substrate by conventional poly(methyl methacrylate) (PMMA) transfer method.[38–40] Sample cleaning step of 300–350 °C *in situ* annealing in preparation chamber (base pressure of 10$^{-10}$ mbar) was done to get rid of unwanted carbon based contamination including residual PMMA. F6TCNNQ (Novaled), as molecular acceptor, was deposited onto the clean MoS$_2$/HOPG in the same preparation chamber while monitoring the nominal mass-thickness by using a quartz crystal microbalance.

**Photoemission measurements.** ARPES spectra were measured at the beamline BL7U (UVSOR, Japan), the IMS (MBS-A1 lab system, Japan) and the Humboldt University (lab system, Germany) using a hemispherical electron analyzer with monochromatic light source of 21 eV and 9 eV, respectively. The determination of total resolution of 100 meV at 300 K were performed using clean Au (111) single crystal. The energy calibration with respect to Fermi level were carried out by measuring the electrically grounded clean Au (111) single crystal.

**Theoretical modeling.** The calculations presented here were based on density functional theory (DFT),[41,42] as implemented in the all-electron, numeric atom-centered electronic-structure package FHI-aims.[43,44] The Perdew-Burke-Ernzerhof (PBE) generalized gradient approximation[45] functional was employed for the atomic geometry relaxation, while the hybrid Heyd-Scuseria-Ernzerhof (HSE06) was used for the electronic properties to grasp the correct charge transfer and level alignment.[46] The "tight" basis sets were used for PBE calculations, while the "intermediate" settings were used for HSE06. All the calculations include the Tkatchenko-Scheffer van der Waals (TS-vdW) interactions[47] and non-self-consistent spin-orbital coupling (SOC).[48] Comparison of TS-vdW with the many body dispersion of Ref. [49] shows that the corrugation of the potential energy surface is similarly described by these methods (details in Supplementary Table S2). A vacuum region of 100 Å and a dipole correction were employed to avoid the spurious electrostatic interactions between periodic supercells.[47] A 20×20×1 k-point sampling was used for the (1×1) ML-MoS$_2$ (0001) and graphite (001) surface to attain the correct electronic properties. Four layers of graphite (001) surface were considered to guarantee the convergence of work function, which is also reported in a previous experiment.[50] The atoms in the two bottom layers were fixed while other atoms are allowed to fully relax. The minimum force on atoms is guaranteed to be below 0.001 eV/Å.

We initially performed a grid search to find the most stable geometry of one molecule adsorbed on free-standing MoS$_2$ with lying-down and standing-up geometries at different packing densities. At the low packing density under the supercell of (6×6), 41 models for lying-down were simulated by PBE functionals, and 10 models were selected to perform the electronic property calculations. For the higher packing densities, 22 models were considered. This method is described in detail in the previous work.[25] To simulate MoS$_2$ (0001)/graphite (001), a typical commensurate superstructure with (4×4)

MoS$_2$ and (5×5) graphite was considered, to reduce the lattice mismatch. Based on this moiré superstructure, 6 models with one molecule lying-down on the supercell graphite (5×10)/MoS$_2$ (4×8) and 3 models with molecule lying-down, short-tilted, and long-tilted on the supercell graphite (5×5)/MoS$_2$ (4×4) were considered to perform the geometry optimization with the PBE functional. We then used a random structure search strategy[51] to find geometries of two molecules lying flat at the (4×8) MoS$_2$ supercells. The most stable structure was selected for consideration with the graphite substrate. In the theory part of this paper, HOPG refers to a graphite (001) (5×10) supercell. All calculations have been performed with a Gaussian occupation smearing with a width of 150 meV to ensure convergence of the self-consistent cycles. The (P)DOS shown in the manuscript have been shifted such that this occupation function assumes values below 0.09 above the zero of energy.

The Hessian matrix was constructed and diagonalized for the F6TCNNQ/ML-MoS$_2$ (4×8)/graphite (5×10) system by using the phonopy package,[52] in which only atoms of F6TCNNQ and MoS$_2$ were displaced by 0.01 Å, in the presence of the graphite substrate. Here, forces were calculated with the PBE+TS-vdW functional in the supercell. The phonon modes obtained in this way were used to create the thermal displacements as detailed in the Section 8 in the Supplemental Information. We then used 35, 50 and 35 configurations to calculate TDOS and PDOS of the F6TCNNQ/ML-MoS$_2$(4×8)/graphite (5×10) [522 atoms] at 50 K, 150 K, and 300 K, respectively.

**Acknowledgments**

This work was funded by the Deutsche Forschungsgemeinschaft (DFG) - Projektnummer 182087777 - SFB 951, AM 419/1-1, and by the JSPS KAKENHI Grant No. JP18H03904. Further support by the National Research Foundation (NRF) of Korea under Grant 2018M3D1A1058793 and Technology Innovation Program (20012502), funded by the Korean Ministry of Trade, industry & Energy, is acknowledged. We thank the IMS and HZB for allocating synchrotron radiation beam time (UVSOR, BL7U and Bessy II, PM4). H. W. thanks Karen Fidanyan for assistance with the phonon calculations.

**Author contributions**

S.P., N.K., and M.R. conceived and supervised the project. S.P., T.S., M.M., T.Y., Y. H., P.A., and S.K. performed ARPES measurements and analyzed the data, under supervision of N.K. A.A., A.H., L.L., S.K., and V.C.T prepared ML-$MoS_2$ samples. H.W., M.Z, D.M., and M.R. performed all calculations and their analysis. S.P., H.W., N.K., and M.R. prepared the manuscript. All authors commented on the manuscript.

**Additional information**

Supporting Information is available from the Wiley Online Library or from the author.

Supplementary Information for

# Temperature-dependent electronic ground state charge transfer in van der Waals heterostructures

Soohyung Park,[1,†] Haiyuan Wang,[2,3,†], Thorsten Schultz,[4,5] Dongguen Shin,[4] Ruslan Ovsyannikov,[5] Marios Zacharias,[2,6] Dmitrii Maksimov,[2,7] Matthias Meissner,[8] Yuri Hasegawa,[8] Takuma Yamaguchi,[8] Satoshi Kera,[8] Areej Aljarb,[9] Mariam Hakami,[9] Lain-Jong Li,[9,10] Vincent Tung,[9] Patrick Amsalem,[4] Mariana Rossi,[2,7,*] and Norbert Koch[4,5,*]

[1] Advanced Analysis Center, Korea Institute of Science and Technology (KIST), Seoul 02792, South Korea

[2] Fritz Haber Institute of the Max Planck Society, 14195 Berlin, Germany

[3] Chaire de simulation à l'échelle atomique (CSEA), Ecole Polytechnique Fédérale de Lausanne (EPFL), CH-1015 Lausanne, Switzerland

[4] Humboldt-Universität zu Berlin, Institut für Physik & IRIS Adlershof, 12489 Berlin, Germany

[5] Helmholtz-Zentrum für Materialien und Energie GmbH, 12489 Berlin, Germany

[6] Department of Mechanical and Materials Science Engineering, Cyprus University of Technology, 3603 Limassol, Cyprus

[7] Max Planck Institute for the Structure and Dynamics of Matter, 22761 Hamburg, Germany

[8] Institute for Molecular Science, 444-8787 Okazaki, Japan

[9] Physical Sciences and Engineering, King Abdullah University of Science and Technology, Thuwal 23955-6900, Saudi Arabia

[10] Department of Mechanical Engineering, The University of Hong Kong, Pok Fu Lam Road, Hong Kong

[†] both authors contributed equally to this work

[*] corresponding authors: rossi@fhi-berlin.mpg.de, norbert.koch@physik.hu-berlin.de

**Table of contents**

**Section 1. Quality of ML-MoS$_2$/HOPG**

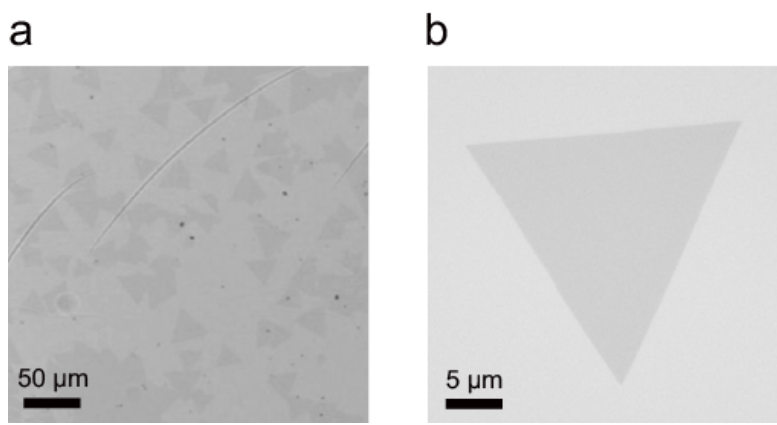

**Supplementary Figure S1** | Optical microscopy images of ML-MoS$_2$/HOPG. (a) Low magnification image from the sample edge with about 50 % coverage of ML-MoS$_2$ on top of HOPG; the sample center (where ARPES data were obtained) has an average ML-MoS2 coverage above 80%. (b) Zoon-in image showing a sharp and equilateral ML-MoS$_2$ triangle.

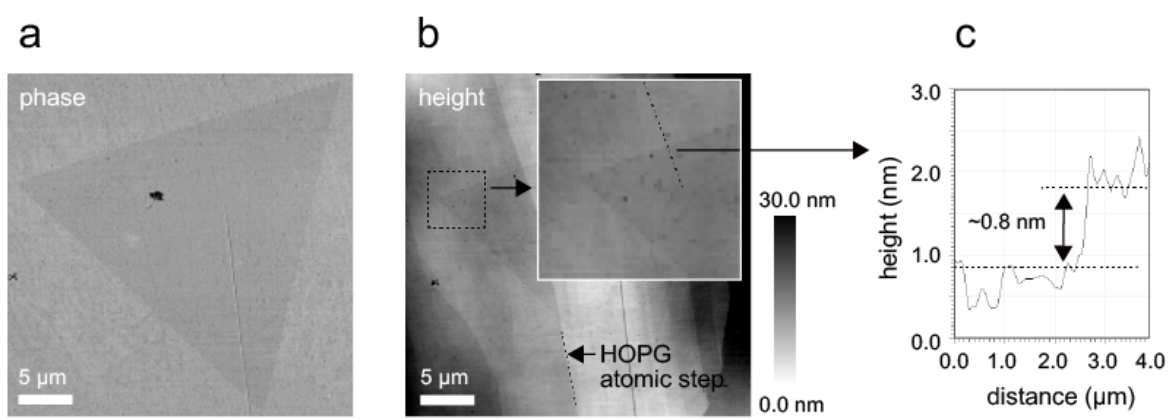

**Supplementary Figure S2** | Atomic force microscopy (AFM) images of ML-MoS$_2$/HOPG. (a) Phase and (b) height AFM image of ML-MoS$_2$. (c) Height profile measured along the dashed line in the inset of (b) indicates ML-height (~ 0.8 nm).

**Section 2. Temperature-independent charge transfer in F6TCNNQ/HOPG**

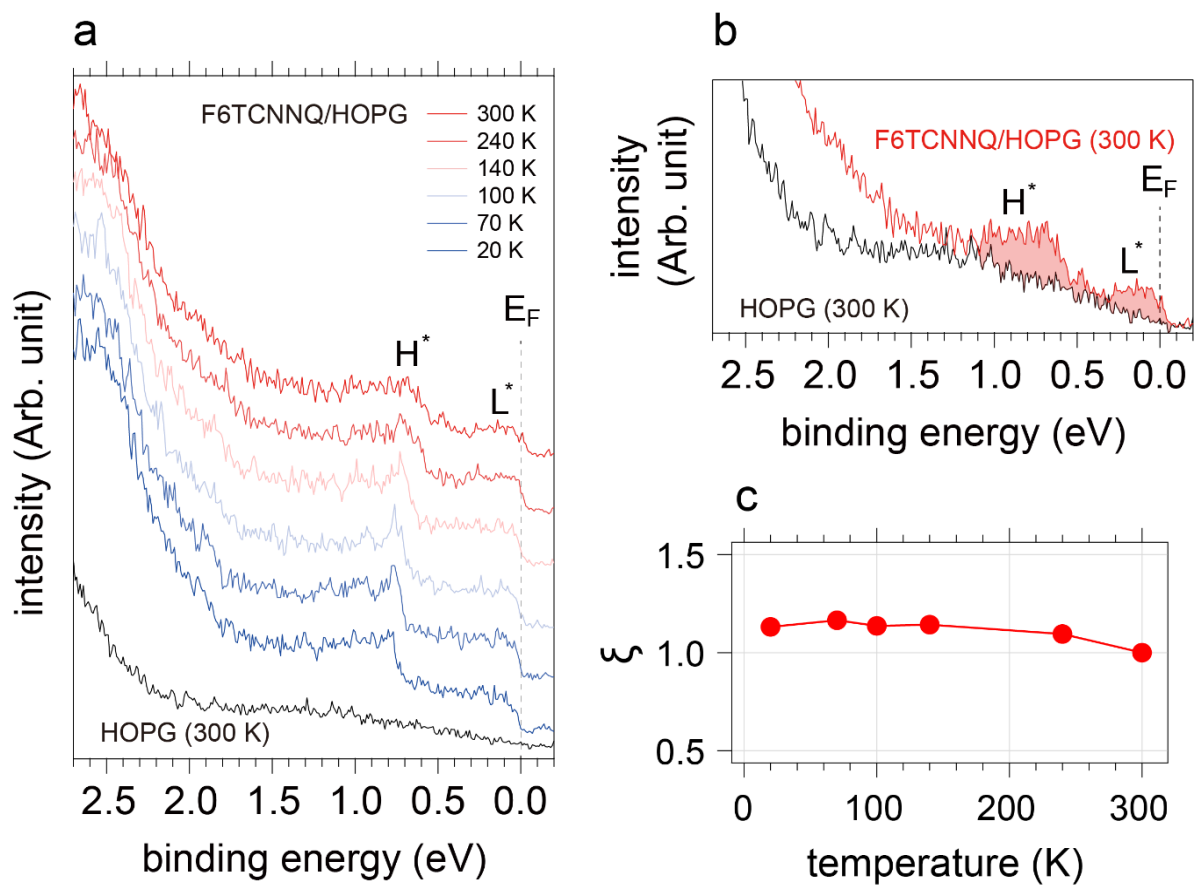

**Supplementary Figure S3 |** F6TCNNQ on HOPG at different temperatures. (a) Energy distribution curves (EDC) of 0.5 monolayer (ML) F6TCNNQ on HOPG and clean HOPG near the Fermi-level as a function of temperature around K point (angle integrated for 22°) by using He I$_\alpha$. (b) EDCs of 0.5 ML F6TCNNQ/HOPG and clean HOPG spectra at 300 K. L* and H* denote the partially filled LUMO and relaxed HOMO of F6TCNNQ representing the charged F6TCNNQ. Colored area indicating the charged F6TCNNQ features, which are summarized in Figure c. (c) Relative change in charged F6TCNNQ ($\xi$) as a function of temperature, which shows that amount of charged F6TCNNQ is not significantly changed by temperature.

To understand the impact of temperature on the charge transfer between F6TCNNQ and HOPG alone, ARPES measurement were carried out as shown in Figure S3 and it shows only a sharpening of the F6TCNNQ features without a change of spectral weight (area). This shows that the ratio of neutral and charged F6TCNNQ molecules is not influenced by temperature.

**Section 3. Estimation of the fraction of charged F6TCNNQ at 300 K.**

To quantitatively estimate the amount of charge transfer, the fraction of charged F6TCNNQ molecules is calculated from the transferred electron density divided by the density of F6TCNNQ on top of the monolayer (ML) MoS$_2$ surface ($\rho_e/\rho_{F6TCNNQ}$), based on the assumption that one electron can be transferred into only one molecule.

(1) Density of F6TCNNQ on top of ML-MoS$_2$ surface

In the F6TCNNQ deposition step, the nominal thickness was monitored by a quartz crystal microbalance (QCM), which enables us to calculate the density of F6TCNNQ on top of ML-MoS$_2$ using the following equation:

$$\rho_{F6TCNNQ} = \frac{N_{F6TCNNQ}}{A} = \frac{S}{\mu}d$$

where $\rho_{F6TCNNQ}$, $N_{F6TCNNQ}$, $A$, $S$, $\mu$ and $d$ are the density of F6TCNNQ on top of ML-MoS$_2$ surface, the number of F6TCNNQ molecules, area, mass density used in monitoring the nominal thickness by QCM, molecular molar mass, and nominal thickness taken from the QCM, respectively. The $S$ of 1.3 g/cm$^3$, $d$ of 5Å (~0.5 ML), and $\mu$ of 362.2 g/mol were used, and the estimated $\rho_{F6TCNNQ}$ of $1.08 \times 10^{14}$ molecule/cm$^2$ was obtained.

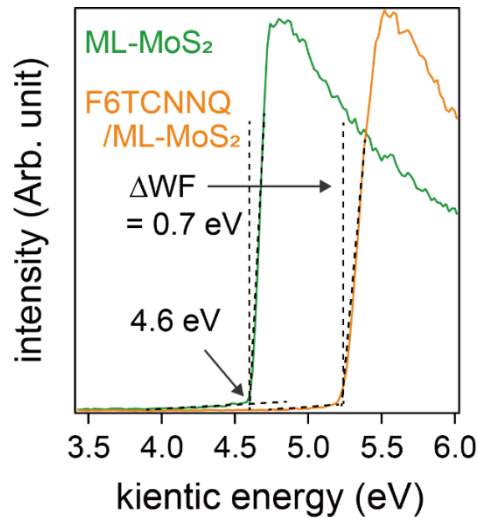

**Supplementary Figure S4** | UPS measurement of secondary electron cutoff region of ML-MoS$_2$ with/without F6TCNNQ at 300 K. The spectrum shift of 0.7 eV shows the change in work function due to charge transfer from HOPG to F6TCNNQ.

(2) Transferred electron density

The density of transferred charges can be calculated using the Helmholtz equation as follows[1]:

$$\rho_e = \frac{\Delta \text{WF} \cdot \varepsilon_{\text{eff}} \varepsilon_0}{e d_{\text{eff}}}$$

where $\Delta \text{WF}, \varepsilon_0, \varepsilon_{\text{eff}}, d_{\text{eff}}$, and $e$ are the work function change, the vacuum permittivity, the effective dielectric constant, the effective dipole distance (HOPG to F6TCNNQ), and the elementary charge, respectively. The parameters $\varepsilon_{\text{eff}}$ = 5.46 (ML-MoS$_2$ for 6.4, F6TCNNQ for 3, and HOPG for 7), $d_{\text{eff}}$ = 9 Å, and the measured $\Delta \text{WF}$ of 0.70 eV were used. As a result, the transferred charge density $\rho_e$ at 300 K was calculated to be $2.35 \times 10^{13}/\text{cm}^2$.

From that, we now can estimate the fraction of charged F6TCNNQ at 300 K to be 21.71 % ($\rho_e/\rho_{\text{F6TCNNQ}}$), and that at 7 K to be 7.24 % in the proportional way using ξ at 7 K as defined in the main manuscript text (ξ$_{@7K}$ ·$\rho_e/\rho_{\text{F6TCNNQ}}$).

**Supplementary Table S1** | Impact of temperature on the amount of charge transfer and the fraction of charged molecules, obtained according to the procedure described above.

| temperature (K) | 300 | 240 | 200 | 140 | 100 | 70 | 7 |
|---|---|---|---|---|---|---|---|
| amount of charge transfer ($\times 10^{13}/\text{cm}^2$) | 2.35 | 2.26 | 2.35 | 2.09 | 1.76 | 1.18 | 0.84 |
| fraction of charged F6TCNNQ molecules (%) | 21.7 | 20.8 | 21.7 | 19.3 | 16.3 | 10.9 | 7.24 |

**Section 4. ARPES spectra of ML-MoS$_2$ + molecular acceptor for different temperatures**

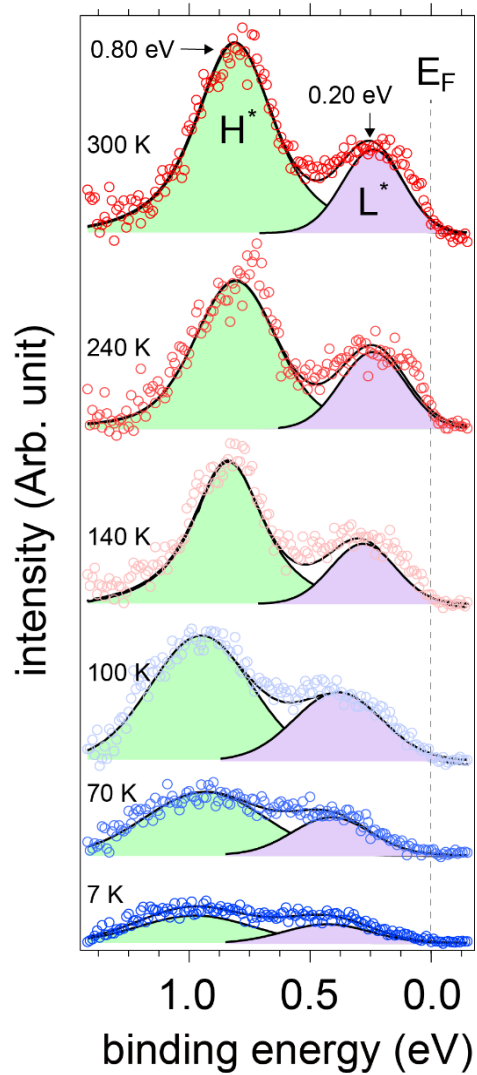

**Supplementary Figure S5 |** Energy distribution curve (EDC) of charged F6TCNNQ as a function of temperature. To obtain the reliable binding energy of filled LUMO (L*) and relaxed HOMO (H*), the HOPG substrate feature were subtracted. The energetic position difference and intensity ratio between H* and L* were fixed and checked from previous work[1] during the fitting process.

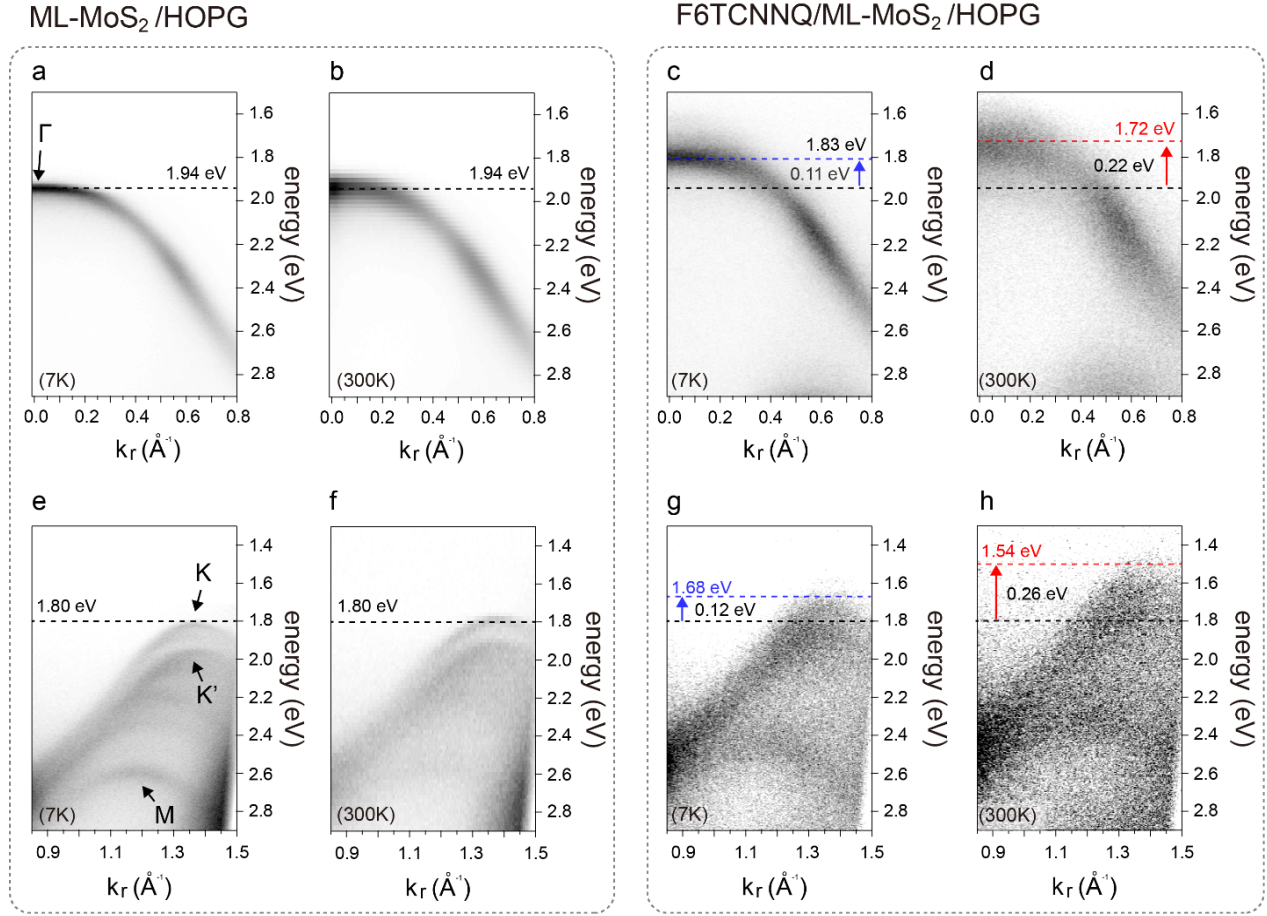

**Supplementary Figure S6 |** Angle resolved photoelectron spectroscopy (ARPES) spectra of ML-MoS$_2$ by molecular doping. ML-MoS$_2$ /highly oriented pyrolytic graphite (HOPG) ARPES spectra were measured around Γ (a-d) and K (e-h) point of the Brillouin zone (BZ) along the Γ to K+M direction. The spectra were collected at 7 K (a, c, e and g) and 300 K (b, d, f and h) with monochromatic photon energy of 21 eV. (c), (d), (g) and (h) corresponding ARPES spectra after deposition of nominally 0.5 ML of F6TCNNQ.

In order to address the effect of temperature on the band structure of ML-MoS$_2$ without F6TCNNQ, ARPES spectra of clean ML-MoS$_2$/HOPG were measured with increasing temperature from 7 K to 300 K in Figure S6 left dashed line frame. Figure S6a and b show that ARPES spectra of ML-MoS$_2$ around the Γ point of BZ for the selected temperatures. The local valence band maximum (VBM) at Γ point is estimated by 1.94 eV, which is in good agreement with previous reported values.[1,2] In Figure S6e and f, we found two split bands with local valley (K-K') around K point due to the lack of inversion symmetry and spin-orbit coupling.[3,4] The overlap of the bands along the two high symmetry directions Γ-M and Γ-K is an intrinsic feature of 2H phase ML-MoS$_2$ in azimuthally disordered sample.[5]

It is noteworthy that the spectra of the two extreme temperatures (7 K, 300 K) are nearly identical except for the spectral broadening originating from thermal fluctuation of the atoms. From that, it can be concluded that the band structure of ML-MoS$_2$ itself does not change by thermal fluctuations within the experimental resolution.

Upon deposition of a nominal 0.5 ML of F6TCNNQ, the band structure of ML-MoS$_2$ moves closer to the Fermi level by ca. 0.11 eV at both Γ and K points of BZ at 7 K as shown in Figure S6c and g. With increasing temperature up to 300 K, the band structure of ML-MoS$_2$ is gradually shifted towards the Fermi level resulting in a lower binding energy of VBM of 1.72 eV at Γ and 1.54 eV at K point, respectively. Therefore, all measured spectra show a gradual and rigid shift toward the Fermi level in the whole BZ upon stepwise increase of the temperature as summarized in Figure 2c in main manuscript. In addition, we also found a significant broadening of spectra originating from inelastic scattering of emitted photoelectrons from ML-MoS$_2$ by the adsorbed molecules.

In short, we did not find any evidence for any change in the electronic structure of ML-MoS$_2$ itself as a function of temperature. On the other hand, it is observed that the amount of ML-MoS$_2$ energy shift when molecules are adsorbed depends notably on temperature.

**Section 5. Determination of orientation of F6TCNNQ on top of ML-MoS$_2$ with temperature**

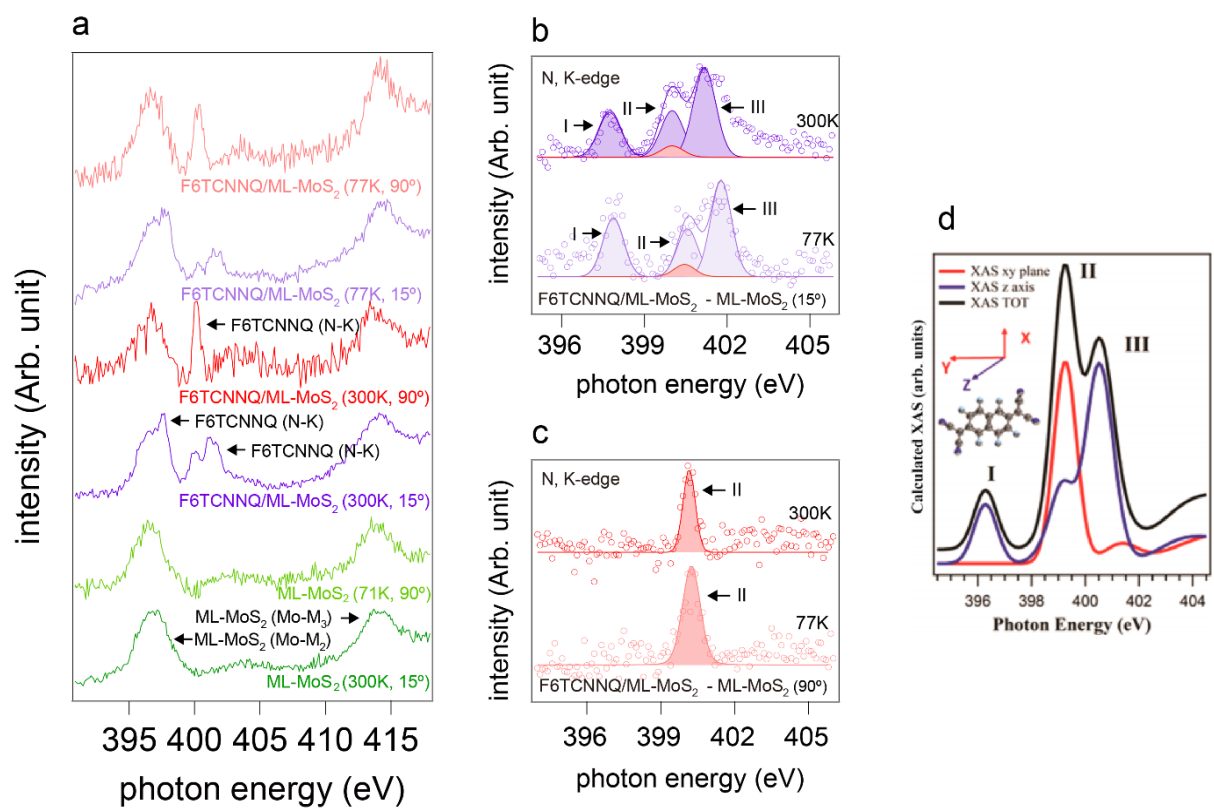

**Supplementary Figure S7** | Temperature dependent x-ray absorption spectroscopy (XAS). (a) Selected XAS spectra of ML-MoS$_2$/HOPG and F6TCNNQ (0.5 ML)/ML-MoS$_2$/HOPG by varying the temperature and the angle to surface. The ML-MoS$_2$ spectra of other conditions are not shown here as they did not exhibit significant changes within the experimental resolution. The arrows denote the XAS edge for Mo M$_2$-edge, M$_3$-edge, and N K-edge, respectively. The ML-MoS$_2$/HOPG-subtracted XAS spectra of F6TCNNQ (0.5 ML)/ML-MoS$_2$/HOPG recorded at (b) 15° and (c) 90° varying the temperature. (d) Simulated *xy* plane (red) and *z*-axis (blue) N K-edge XAS spectra of F6TCNNQ. The xy plane and z-axis spectra denote that the absorption in the xy plane of F6TCNNQ and axis perpendicular to F6TCNNQ, respectively. XAS TOT (black) is the simulated total absorption. Reprinted with permission from Ref. 6. Copyright (2015) American Chemical Society

In order to determine the orientation of F6TCNNQ on top of HOPG at different temperatures, x-ray absorption spectroscopy was performed at the beamline PM4 (Bessy II, Germany). First, ML-MoS$_2$/HOPG was annealed at 300°C in the preparation-chamber to obtain a clean surface. The XAS spectra of clean ML-MoS$_2$/HOPG in N-K edge energy range were measured by varying the temperature from 300 K to 71 K and the angle between incident beam and surface normal from 15° to 90°, respectively. The spectra of ML-MoS$_2$/HOPG measured at different angles and temperatures are identical, therefore, only two selected extreme cases are shown in Figure S7 bottom two spectra. Since there were no nitrogen atoms present, only the Mo-$M_2$ and $M_3$ edge features originating from ML-$MoS_2$ are observed without temperature and angle dependency. This enables to subtract the overlapping of ML-$MoS_2$/HOPG features easily from the spectra of F6TCNNQ/ML-$MoS_2$/HOPG as shown in Figure S7b and explained below.

As shown in the middle two spectra in Figure S7, varying the angle between the sample and incident light significantly changes the N-K edge feature coming from the F6TCNNQ. On the other hand, the change of temperature does not impact the N-K edge feature except for its sharpness. This is a clear evidence that F6TCNNQ on ML-$MoS_2$ adopts the same orientation with respect to the surface normal regardless of temperature.

As plotted in Figure S7b and c, we then subtract the contribution from ML-$MoS_2$ from the XAS spectra of F6TCNNQ/ML-$MoS_2$/HOPG in order to further analyze the N-K edge of F6TCNNQ. In Figure S7b and c, XAS spectra are further deconvoluted by one red and three violet Gaussians, according to previous reports as shown in Figure S7d (xy-plane for red peak and z-axis consisting of three blue peaks), respectively.[6] From this assignment, the orientation of F6TCNNQ is found to be lying on top of ML-$MoS_2$ surface regardless of the temperature.

**Section 6. Calculated properties of F6TCNNQ on the free-standing ML-MoS₂**

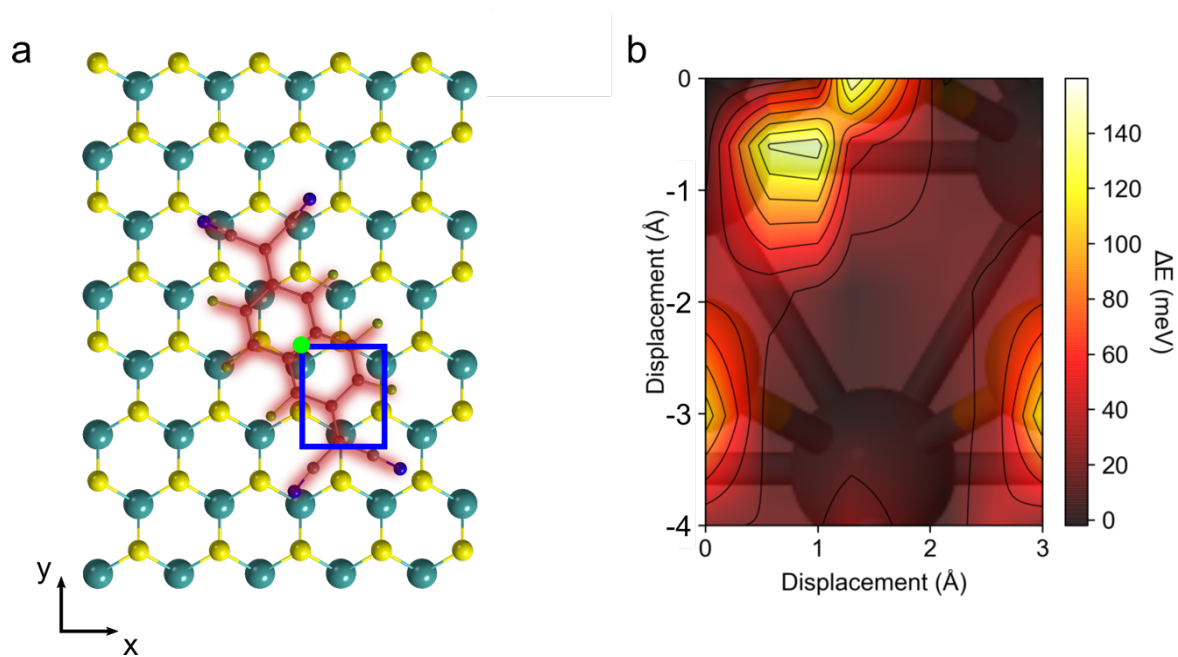

**Supplementary Figure S8** | The atomic structure of F6TCNNQ on ML-MoS$_2$ (a) Lowest energy structure of F6TCNNQ on ML-MoS$_2$ as obtained with the HSE06+TS-vdW functional. (b) Corrugation of the potential energy landscape at different adsorption sites of F6TCNNQ on ML-MoS$_2$. The energy landscape was obtained by shifting the centre of mass of the molecule, constraining the central 2 carbon atoms in the x and y direction and relaxing all other degrees of freedom. The axis denotes the position marked by the green point in Figure a, and the area in the blue box corresponds to Figure b.

**Supplementary Table S2 | Impact of vdW interactions.** For four adsorption structure models randomly selected from the free standing F6TCNNQ/ML-MoS$_2$ structures and the most stable F6TCNNQ/ML-MoS$_2$/HOPG model MBD-nl calculations were performed. The adsorption energy per molecule is defined as $E_{\text{ads}} = \frac{1}{n}\left(E^{\text{total}}_{\text{F6TCNNQ/MoS}_2\text{/HOPG}} - E^{\text{total}}_{\text{F6TCNNQ}} - E^{\text{total}}_{\text{MoS}_2\text{/HOPG}}\right)$, where $n$ is the number of molecule, and $E$ is the total energy of different slabs.) All the results based on fully relaxed geometries from HSE06+TS here. TS overestimates the adsorption energy but predicts a similar corrugation of the energy landscape as MBD-nl (see energy difference between different models).

|  |  | $E_{\text{ads}}$ (eV) | |
|---|---|---|---|
|  |  | TS | MBD_nl |
| F6TCNNQ/ML-MoS$_2$ | Model 1 | -1.764 | -1.101 |
|  | Model 2 | -1.825 | -1.156 |
|  | Model 3 | -1.849 | -1.166 |
|  | Model 4 | -1.828 | -1.144 |
| F6TCNNQ/ML-MoS$_2$/HOPG |  | -1.514 | -0.773 |

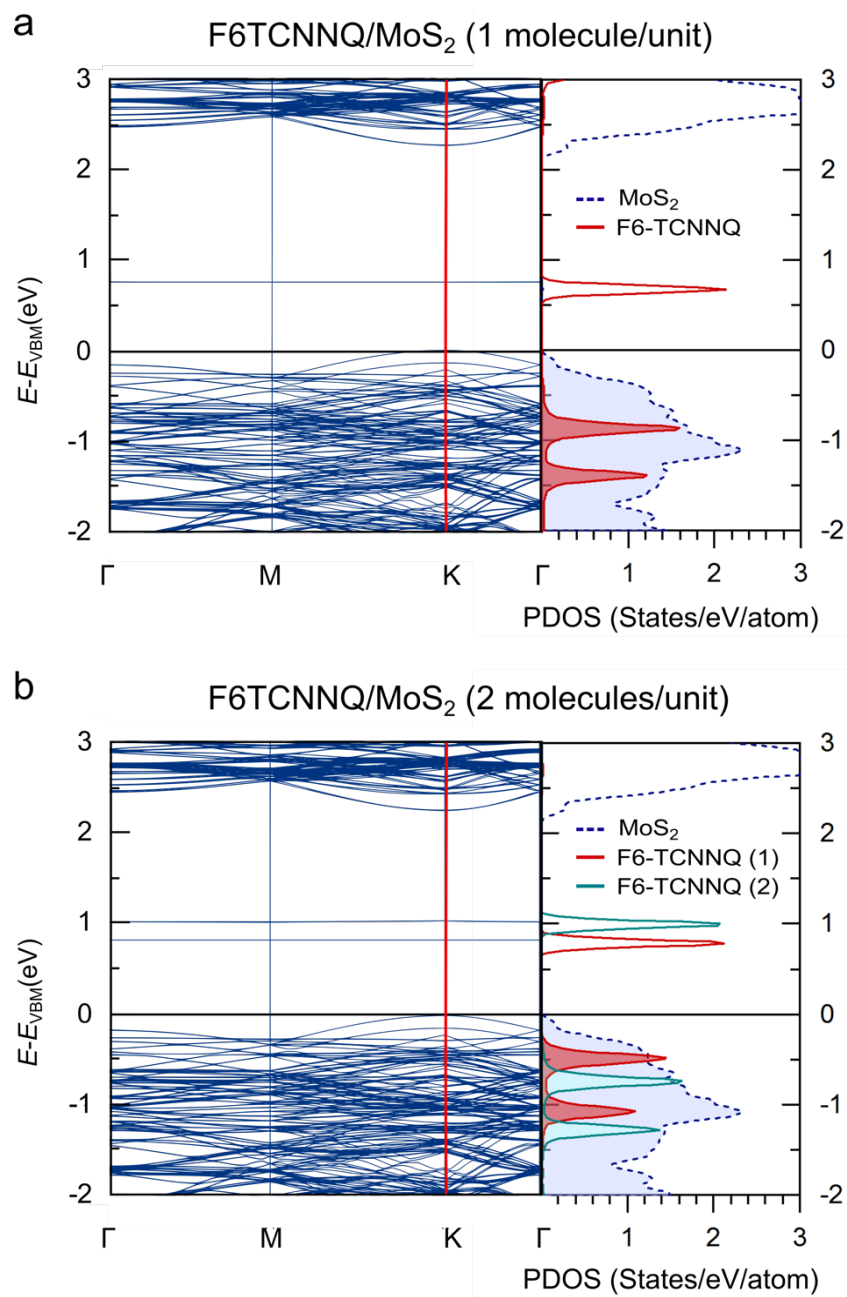

**Supplementary Figure S9** | Calculated electronic band structure and projected density of states. Calculated band structure and projected density of states for dilute (a) and ML (b) F6TCNNQ on free-standing ML-MoS$_2$ (4×8 supercell) with the HSE06 functional. Band structure is folded.

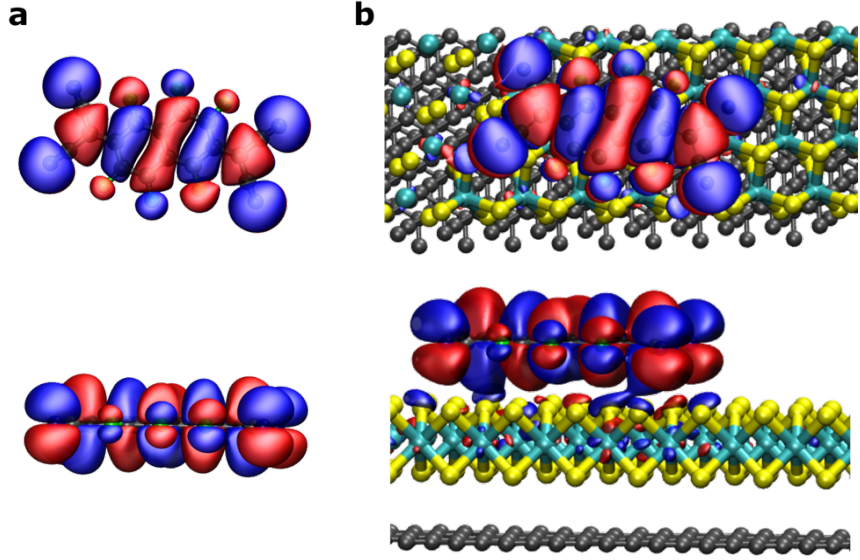

**Supplementary Figure S10 |** Visualization of electronic orbitals in F6TCNNQ/ML-MoS$_2$/HOPG. Visualization of (a) the LUMO of the isolated F6TCNNQ molecule and (b) of the partially filled LUMO in the F6TCNNQ/ML-MoS$_2$/HOPG system. Isosurfaces of +/- 0.01 e/Å$^{3/2}$ are shown.

**Section 7. Temperature dependence of electronic levels**

We here address temperature-dependence, caused by electron-phonon coupling, of the electronic density of states obtained from the single-particle Kohn-Sham levels $\varepsilon_{n,k}$ calculated with the HSE06 functional. Following the discussion in Ref. 7, we separate this temperature dependence as follows. Assume that a variation in the electronic density of states $D(E)$ as a function of volume $V$ and temperature $T$ is given by

$$dD(E,V,T) = \left(\frac{\partial D}{\partial T}\right)_V dT + \left(\frac{\partial D}{\partial V}\right)_T dV. \qquad (1)$$

Then we can write that its variation with temperature at a constant pressure is

$$\left(\frac{\partial D}{\partial T}\right)_P = \left(\frac{\partial D}{\partial T}\right)_V + V\left(\frac{\partial D}{\partial V}\right)_T \frac{1}{V}\left(\frac{\partial V}{\partial T}\right)_P = \left(\frac{\partial D}{\partial T}\right)_V + \left(\frac{\partial D}{\partial \ln V}\right)_T \left(\frac{\partial \ln V}{\partial T}\right)_P. \qquad (2)$$

In Eq. 2, we identify that the full temperature dependence is composed by the variation of $D$ with temperature at a fixed volume plus its variation with volume at a fixed temperature, multiplied by the thermal expansion coefficient $\alpha = \left(\frac{\partial \ln V}{\partial T}\right)_P$.

We address the last term of Eq. 2 by taking the experimental thermal expansion coefficients for graphite and ML-MoS$_2$ from references[8,9] and scaling the ground-state lattice constants by this factor at the temperatures of 50, 150 and 300 K. Since there is a mismatch in the in-plane expansion coefficient of graphite and ML-MoS$_2$, we approximated these lattice constants by their predicted average at each temperature, in order to minimize strain. We then calculated the ground-state $D(E)$ at each volume, consistent with $T$ = 50, 150, 300 K, which is shown in Figure S11. The first term of Eq. 2 describing the electron-phonon effect on $D$, is more challenging to address, and we detail the procedure in the following section.

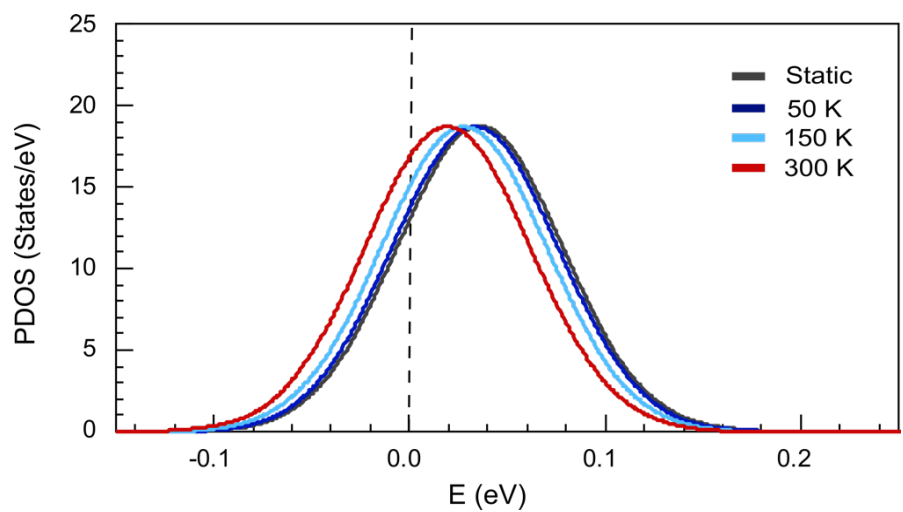

**Supplementary Figure S11** | The PDOS of F6TCNNQ in the system considering only lattice expansion at different temperatures. Projected density of states has been consistently shifted by 5.4 eV, see main text.

## Section 8. Stochastic sampling of the vibrational space

In the framework of the Williams-Lax theory[10,11] and harmonic approximation, the temperature dependence of the electronic density of states $D$ is obtained as a multidimensional Gaussian integral[12,13]:

$$D(E,T) = \prod_\nu \int dQ_\nu \frac{e^{-Q_\nu^2/2\sigma_{\nu,T}^2}}{\sqrt{2\pi}\sigma_{\nu,T}} D(E,Q), \quad (3)$$

where the product runs over all modes $\nu$, and $Q$ is used to indicate the configuration defined by the normal coordinates $\{Q_\nu\}$. $D$ evaluated at the configuration $Q$ is given by:

$$D(E,Q) = \sum_{n\mathbf{k}} \omega_\mathbf{k} \delta(E - \varepsilon_{n\mathbf{k}}(Q)), \quad (4)$$

where the summation is taken over all bands $n$ and wave vectors $\mathbf{k}$ with weights $\omega_\mathbf{k}$, and $\varepsilon_{n\mathbf{k}}(Q)$ represents the Kohn-Sham energies of the system evaluated at configuration $Q$. Equation 3 represents precisely the thermal average of $D$ at the fixed volume, and can be numerically evaluated using importance sampling Monte Carlo integration. This corresponds to (i) generating a variety of atomic configurations $Q$, at the fixed volume, from the distribution $\prod_\nu e^{-Q_\nu^2/2\sigma_{\nu,T}^2}/\sqrt{2\pi}\sigma_{\nu,T}$, (ii) calculating the $D(E,Q)$ for each configuration, and (iii) taking the average of all calculated $D(E,Q)$. Regarding step (i), the widths of the multivariate Gaussian distribution are defined by the mean square displacement of the atoms along mode $\nu$ as[14]:

$$\sigma_{\nu,T} = l_\nu \sqrt{2n_{\nu,T} + 1}, \quad (5)$$

where $l_\nu = \sqrt{\hbar/2M_p\omega_\nu}$ is the zero-point vibrational amplitude, and $n_{\nu,T} = [e^{\frac{\hbar\omega_\nu}{k_B T}} - 1]^{-1}$ is the Bose-Einstein distribution. Additionally, $M_p$ and $\omega_\nu$ denote the mass of proton and the frequency of the $\nu$th normal mode. Taking Eq. 5 together with the set of frequencies and eigenmodes obtained with phonopy,[15] we can create a list of atomic displacements at a given temperature. In the following, the method of Sobol low-discrepancy sequences[16] was used to sample efficiently the normal coordinates.[12] Finally, 35, 50 and 35 configurations were used to calculate the electronic density of states (EDOS) of the F6TCNNQ/ML-MoS$_2$(4×8)/graphite (5×10) [522 atoms] for temperatures 50 K, 150 K, and 300 K, respectively. This amount of sampling ensured sufficient statistical convergence for the effects discussed in this manuscript. In the same way with the density of sates in the Eq. 3, we also calculated the temperature-dependent band gap for the free-standing ML-MoS$_2$ presented in Figure S12 and Table S3. The band gap becomes smaller as the temperature increases, with VBM and CBM being shifted up and down in energy, respectively, and the shift we observe agrees with the literature.[13,17] We have also calculated the renormalization of the levels of the F6TCNNQ molecule as shown in Table S4.

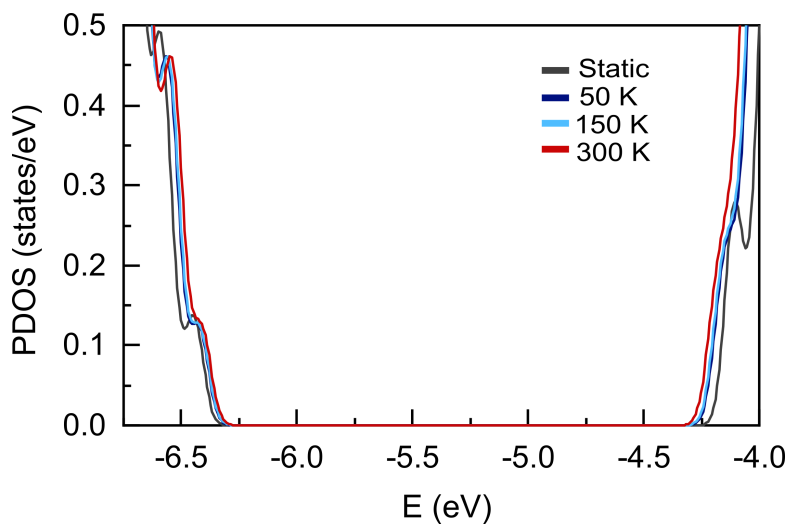

**Supplementary Figure S12** | The effect of thermal displacement on DOS of ML-MoS$_2$. The DOS of ML-MoS$_2$ at different temperatures.

To complement the results shown for the full heterostructure in the main manuscript, in Figures S13(a) and (b) we also show the PDOS of MoS$_2$ and HOPG at 150 K. The hybridization of the MoS$_2$ states with the molecular states is clear. The band gap renormalization of MoS$_2$ is also confirmed, and the depopulation of HOPG states can be observed.

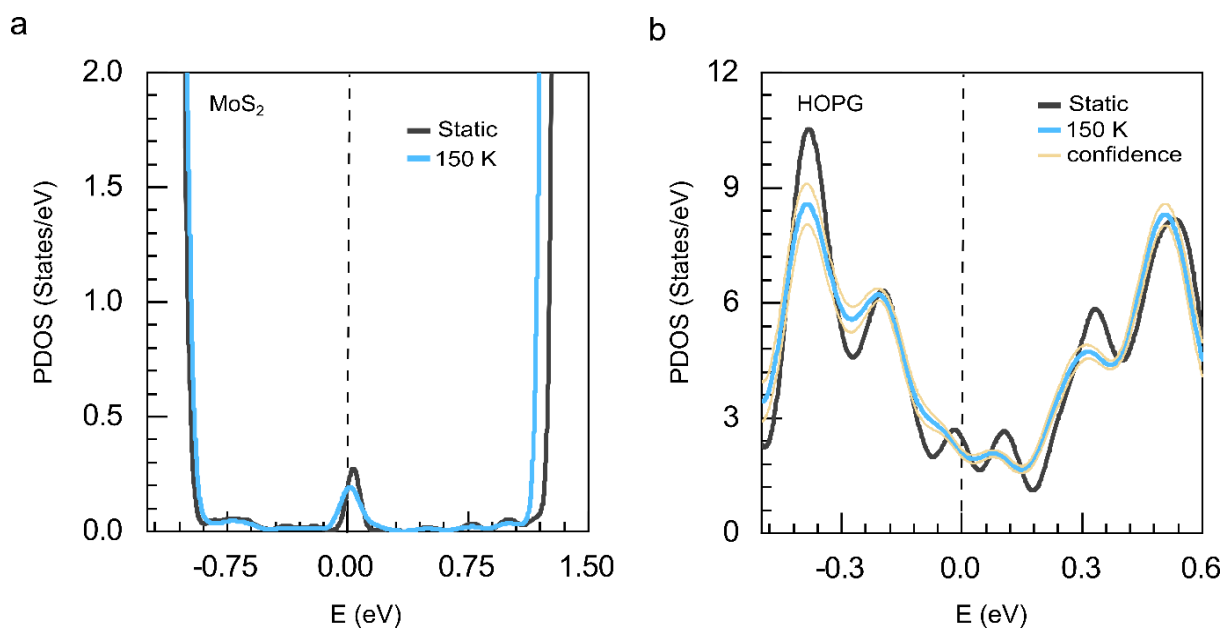

**Supplementary Figure S13** | Projected electronic density of states of MoS$_2$ and HOPG for the static case and at 150 K for F6TCNNQ/ML-MoS$_2$(4×8)/graphite [522 atoms] (HSE06 functional). The curves were all rigidly shifted by 5.4 eV, see main text.

**Supplementary Table S3 |** VBM and CBM renormalization by thermal displacement of ML-MoS$_2$. Position of VBM and CBM of ML-MoS$_2$ as a function of temperature corresponding to Figure S12 (HSE06 functional, ZORA, SOC). In parenthesis, relative difference to static, in meV.

| Temperature (K) | VBM (eV) | CBM (eV) | Band gap (eV) |
| --- | --- | --- | --- |
| Static (0 K) | -6.415 | -4.075 | 2.340 |
| 50 K | -6.394 | -4.117 | 2.277 |
| 150 K | -6.391 | -4.123 | 2.268 |
| 300 K | -6.381 (+34) | -4.143 (-68) | 2.238 (-102) |

**Supplementary Table S4 |** HOMO and LUMO renormalization by thermal displacement of F6TCNNQ. Position of HOMO and LUMO of isolated F6TCNNQ as a function of temperature (HSE06 functional, ZORA, SOC). In parenthesis, relative difference to static, in meV.

| Temperature (K) | HOMO (eV) | LUMO (eV) | Band gap (eV) |
| --- | --- | --- | --- |
| Static (0 K) | -6.886 | -5.903 | 0.983 |
| 300 K | -6.874 (+12) | -5.990 (-87) | 0.884 (-99) |

**Supplementary Table S5 |** Summarized energy level donor-bridge-acceptor at 0 K. Level alignment of isolated components at the potential energy surface (no temperature). All energies in eV. (intermediate for HSE06 and tight basis sets for PBE). Relative alignment in parenthesis.

| Functional | E$_F$ | VBM | CBM | HOMO | LUMO |
| --- | --- | --- | --- | --- | --- |
| HSE06 | -4.737 (0.0) | -6.415 (-1.678) | -4.075 (0.662) | -7.453 (-2.716) | -5.918 (-1.181) |
| PBE | -4.439 (0.0) | -5.875 (-1.436) | -4.146 (0.293) | -6.891 (-2.452) | -5.909 (-1.470) |

**Section 9. Minimal model to understand T dependence of level occupations**

We write the following single-particle donor-bridge-acceptor Hamiltonian in the basis of the isolated systems (commonly called the diabatic basis)

$$H = \epsilon_D |D\rangle\langle D| + \sum_i^2 \epsilon_{B_i}|B_i\rangle\langle B_i| + \epsilon_A |A\rangle\langle A| + \sum_i^2 \delta_{DB_i}|D\rangle\langle B_i| + \sum_j^2 \delta_{AB_j}|A\rangle\langle B_j| + \delta_{DA}|D\rangle\langle A|, \quad (6)$$

where $\epsilon$ refer to the energy levels of the isolated systems and $\delta$ to the diabatic couplings between different levels. Taking the energy $\epsilon_D$ to set the chemical potential in this system, we are interested in knowing, upon interaction, what will be the final population of the A state, i.e., what is the CT from D to A *via* B. Upon solution of this model, the eigenvalues $\epsilon_j$ and eigenvectors $|j\rangle$ of the interacting system are obtained. We are then interested in calculating the population of the original states after interaction. In a single-particle picture, this can be obtained by

$$p_X = \frac{\langle X|e^{-H/k_BT}|X\rangle}{\sum_j e^{-\epsilon_j/k_BT}} = \frac{\sum_j \langle X|j\rangle\langle j|X\rangle}{\sum_j e^{-\epsilon_j/k_BT}}, \quad (7)$$

where $X$ is the state of interest. In this expression, the chemical potential is implicitly assumed to be at zero of energy. One can thus, e.g., solve the model by placing the Fermi level of graphite (D) at 0 eV and expressing the relative position of the other three levels with respect to it, taking their values at the potential energy surface, as shown in Table S5. We assume that $\delta_{DB_i} = 0.5$, $\delta_{AB_i} = 0.2$ and $\delta_{DA_i} = 0.01\ eV$. These are empirical values but reasonable for the vdW bonded systems regarded here and the distances involved between the components.[18] A first-principles calculation of these coupling parameters is desirable but could not be achieved yet for these systems with the electronic-structure codes we employ. We vary $T$ for the along the temperatures considered in this study.

We then let the position of the VBM of ML-MoS$_2$ and the acceptor LUMO vary in the range we have calculated in Tables S3 and S4 (we found that varying the CBM made no difference to the results). The variation of the population of the molecular level is shown in Figure S14. This allows us to exemplarily visualize how the population of acceptor state A (LUMO of F6TCNNQ) depends on the energy $\epsilon_A$ and the energy of the VBM of ML-MoS$_2$ ($\epsilon_{B1}$), which both vary as function of temperature as shown in Tables S3 and S4.

We can conclude the following: (i) The direction of the renormalization of the state energies can lead to an appreciable increase of the molecular state population. This is due to the coupling between the different parts of the system and the electronic level renormalization. (ii) The quantitative amount of variation of the population of the molecular levels will depend on the actual values of the coupling.

However, varying the coupling constants within a sensible range (0.1-0.8 eV) does not change the qualitative picture shown in this paper.

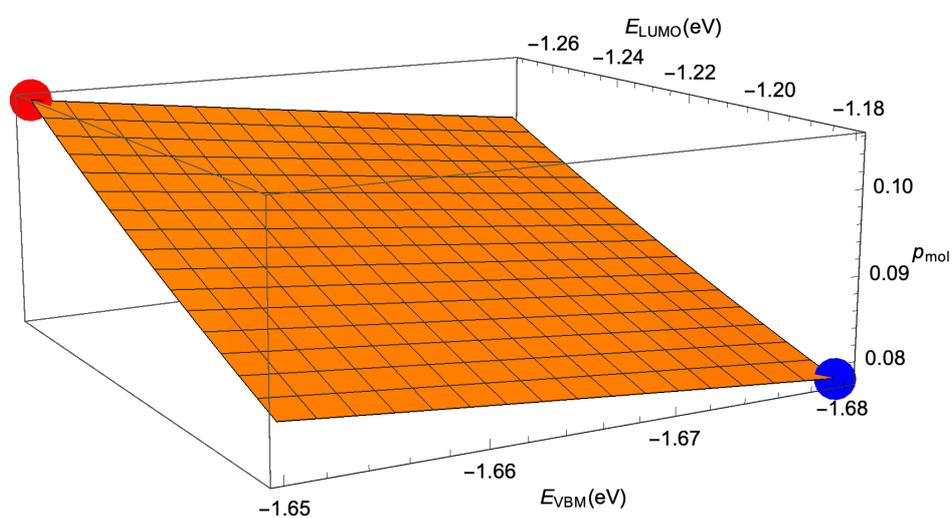

**Supplementary Figure S14** | Population of the molecular level upon a linear variation of the relative VBM and LUMO energies. The blue dot corresponds to the values at the potential energy surface and the red dot corresponds to population with the renormalized levels at 300 K.